\documentclass[preprint2]{aastex}

\def\psrc{PKS~1830$-$211}
\def\calib{J1733$-$1304}

\def\tcmb{\ifmmode{T_{\rm CMB}}\else{$T_{\rm CMB}$}\fi}
\def\tcmbz{\ifmmode{T_{\rm CMB,\,0}}\else{$T_{\rm CMB,\,0}$}\fi}
\def\tex {\ifmmode{T_{\rm ex}}\else{$T_{\rm ex}$}\fi}
\def\trot {\ifmmode{T_{\rm rot}}\else{$T_{\rm rot}$}\fi}
\def\hctn{HC$_3$N}
\def\hctnu{HC$_3$N($3\leftarrow2$)}
\def\hctnk{HC$_3$N($5\leftarrow4$)}
\def\kk{($5\leftarrow4$)}
\def\jfivefour{$J=5\leftarrow4$}
\def\jthreetwo{$J=3\leftarrow2$}
\def\d    {\ifmmode {{\rlap{.}}^\circ}\else {${\rlap{.}}^\circ$}\fi}
\def\s    {\ifmmode {{\rlap{.}}^{\rm s}}\else {${\rlap{.}}^s$}\fi}
\def\as   {\ifmmode {{\rlap{.}}{''}}\else {${\rlap{.}}{''}$}\fi}
\def\kms  {km~s$^{-1}$}
\def\etal {et al.~}
\def\eg   {e.g.,~}
\def\ie   {i.e.,~}

\slugcomment{2012 December 12}
\shorttitle{CMB temperature at Redshift 0.89} 
\shortauthors{Sato \etal}

\begin{document}

\title{On Measuring the CMB Temperature at Redshift 0.89}

\author{M. Sato\altaffilmark{1}, M. J. Reid\altaffilmark{2}, 
 K. M. Menten\altaffilmark{1} and C. L. Carilli\altaffilmark{3} } 
\altaffiltext{1}{Max-Planck-Institut f\"ur Radioastronomie,
 Auf dem H\"ugel 69, 53121 Bonn, Germany}
\altaffiltext{2}{Harvard-Smithsonian Center for Astrophysics,
 60 Garden Street, Cambridge, MA 02138, USA}
\altaffiltext{3}{National Radio Astronomy Observatory,
 Socorro, NM 87801, USA} 

\begin{abstract}
We report on a measurement of  the temperature of the cosmic microwave background
 radiation field, \tcmb, at $z = 0.88582$ by imaging \hctnu\ and \kk\ absorption
 in the foreground galaxy of the gravitationally lens magnified radio source
 \psrc\ using the Very Long Baseline Array and the phased Very Large Array.
Low-resolution imaging of the data
 yields a value of { $\trot = 5.6^{+2.5}_{\,-0.9}$}~K,
{ for the rotational temperature,} \trot, 
 which is { consistent} with the temperature of the cosmic microwave background
 at the absorber's redshift of $2.73 (1+z)$~K.
However,  our high-resolution imaging 
 reveals that the absorption peak position of the foreground gas
 is offset from the continuum peak position of the synchrotron radiation from \psrc~SW,
 which indicates that the absorbing cloud is covering { only part of the emission from} \psrc, rather than the { entire core-jet region}.
 { This changes the line-to-continuum ratios, and we find  
 \trot\ between 1.1 and 2.5 K,}
 which is lower than the expected value.
This shows that previous, \trot, measurements could be { biased}
 due to unresolved structure.

\end{abstract}

%cosmology: cosmic microwave background --- 
 cosmic background radiation --- 
 cosmology: observations --- %gravitational lensing ---
 galaxies: ISM --
 quasars: absorption lines --- quasars: individual (\psrc)
 
\section{INTRODUCTION} 

\psrc\ is a well-studied high-redshift blazar
 ($z = 2.507$; \citealt{Lidman:99}), { which is}
 gravitationally lensed by a foreground face-on spiral galaxy
 revealed by Hubble Space Telescope imaging
  \citep{Wiklind:98, Courbin:02, Winn:02}
 at $z=0.88582$ \citep{Wiklind:96}.
The radio core of \psrc\ is lensed into two components:
 the southwest (SW) and northeast (NE) images
 \citep{PrameshRao:88, Subrahmanyan:90}
  with a separation of 0\as971 \citep{Jin:03}
 or $\simeq$ 7.7~kpc 
{ at $z=0.89$
 (assuming a flat universe with the cosmological parameters
  $H_0 = 70$~\kms~Mpc$^{-1}$, $\Omega_m = 0.27$ and $\Omega_\Lambda = 0.73$).}
The two bright core images are connected
 by a ring-like structure
 of low surface brightness that is often referred to as
 an Einstein ring \citep{Jauncey:91}, whose center is
 near the nucleus of the lensing $z=0.89$ galaxy
  (\citealt{Kochanek:92, Nair:93};
  see \citealt{Courbin:02} for a historic summary).
The SW and NE lensed images of \psrc\ lie on both sides of
 the bulge of the lensing galaxy, with projected distances 
 of $\simeq 3.0$~kpc (0\as40) and $\simeq 4.4$~kpc (0\as59) 
 from the nucleus, respectively \citep{Winn:02, Menten:08}.

Radiation from the SW lensed component of \psrc \ is strongly affected by 
 the intervening gas in the foreground $z=0.89$ galaxy
 that causes reddening and strong absorption features detected
 for a large number of different molecular species
 \citep{Wiklind:96, Wiklind:98, Frye:97, Carilli:98, Menten:99,
  Swift:01, Menten:08, Henkel:08, Henkel:09, Muller:11}.
Although the disk of the $z=0.89$ galaxy is seen
 nearly face-on (inclination $i\lesssim20^\circ$; \eg \citealt{Wiklind:98}),
 the velocity width of the absorption toward the SW image reaches
 $\simeq 100$~\kms\ FWZP \citep{Muller:08}, 
 which is significantly larger than that of most giant molecular clouds (GMC)
 of the Milky Way and better resembles Galactic Center clouds
 \citep{Menten:99, Muller:08}.
Toward the NE lensed image, weaker and narrower
 ($\simeq 15$~\kms\ FWHM) absorption components
 have been detected at a velocity shifted by $-147$~\kms\
 relative to the absorption toward the SW image
 \citep{Wiklind:98, Muller:06}.
The characteristics of the absorption in front of the NE image are
 similar to those of the Galactic diffuse clouds \citep{Muller:08}.

High-redshift molecular absorption provides an important means
 to understand the physical conditions in the molecular interstellar medium
 in a galaxy at a cosmologically significant redshift.
Of particular interest is an accurate measurement of the temperature
 of the cosmic microwave background (CMB)
 radiation (\tcmb),
 one of the fundamental cosmological parameters,
 and its variation with redshift
 (\eg \citealt{Noterdaeme:11}).
The standard cosmological model predicts,
  for an adiabatic expansion of the Universe, 
 $\tcmb (z) = \tcmbz (1+z)$,
 where $\tcmbz=2.73$~K \citep{Mather:94, Mather:99, Fixsen:96}.
This important relation predicted by the theory needs to be confirmed by direct observations.
This can be achieved by measuring the excitation temperature (\tex)
 of interstellar atomic or molecular species at high redshift, 
 in particular the rotational excitation temperature (\trot) of molecular transitions.
In diffuse interstellar gas, where collisional excitation is negligible, 
 rotational transitions can be in radiative equilibrium with the CMB (\ie $\trot = \tcmb$).
The rotational temperature \trot\ of the gas
 can be calculated by observing two transitions of a given molecule,
 and therefore \tcmb\ at the redshift of the gas can be obtained. 
For \psrc, this was done by \citet{Wiklind:96},  who determined an upper limit of
 $\tcmb \le 6.0$~K.
They found that only 36\% of the background continuum
  was covered by the foreground absorbing gas, implying
  observations with higher angular resolution were needed to achieve a reliable value for \trot. 
{ A few years later, \citet{Combes:99} observed a larger sample of molecular transition lines,
 resulting in several measurements of \tcmb\ below the 
 predicted value. 
}
Subsequently, \citet{Henkel:09} performed similar observations with
 the 100-m Effelsberg radio telescope, deriving the covering factor and optical depth
 from advanced models fitted to the data. 
Their analysis was largely consistent with a \tcmb\ value of 5.14 K,
 although individual models could show large offsets. 
Finally, \citet{Muller:11} performed this analysis with
 the Australia Telescope Compact Array, providing a higher resolution. 
Their findings were { mostly} consistent with \tcmb\ = 5.14 K,
 { although some transitions showed significantly lower values than the theoretical prediction.}
Even higher resolution observations are needed to derive a precise measurement.

In this paper, we report on 
 a measurement of  the temperature of the microwave background
 radiation field at $z = 0.88582$ by VLBI imaging of two microwave transitions of \hctn.
The observations and data reduction are described in Section~\ref{sect:obs}.
Our measurement of \trot\ is presented in Section~\ref{sect:results}, using
 both low- and high-resolution imaging (Sections~\ref{sect:lowres} and \ref{sect:hires},
 respectively), followed by a discussion of possible biases in Section~\ref{sect:tint}. 
Finally, our conclusions are summarized in Section~\ref{sect:concl}.

%=========================================================================
\section{OBSERVATIONS AND DATA ANALYSIS}\label{sect:obs}

\subsection{Observations} 

We observed the redshifted \hctnu\ and \hctnk\ absorption lines 
 at 14.5~GHz and 24.1~GHz, respectively,
 toward \psrc\ using the National Radio Astronomy
 Observatory's\footnote[4]{The National Radio Astronomy Observatory
 is a facility of the National Science Foundation operated under
 cooperative agreement by Associated Universities, Inc.}
 Very Long Baseline Array (VLBA)
 combined with the phased Very Large Array (VLA) as a VLBA antenna
 on 1999 January 29 (program BC087B).
We employed two frequency bands of 8 MHz bandwidth each,
 in both right- and left-circular polarizations.
The center frequencies of the two bands were set to cover both velocities
 of the NE and SW components (with $147$~\kms\ difference; \citealt{Wiklind:98}):
respectively,
 14.46272~GHz and 14.47419~GHz for the  \jthreetwo\ transition and
 24.10444~GHz and 24.12356~GHz for the \jfivefour\ transition.
 
We used seven observing blocks of 1 hour duration each.
Subsequent observing blocks were undertaken at alternating frequencies of 14.5~GHz and 24.1~GHz
 (in total, four observing blocks at 14.5~GHz and three at 24.1~GHz).
Each 1 hour block consisted of five repetitions of 12-minute groups, each consisting of 
 a 3-minute scan toward the quasar \calib\ as passband calibrator and
 a 9-minute scan toward the target source \psrc.

The data correlation was performed at the VLBA correlation facility in Socorro, NM.
The data from each antenna pair were cross-correlated with an integration time of 2~seconds.
Each 8 MHz band was split into 128 spectral channels,
 yielding a channel separation of 62.5 kHz.
The corresponding velocity resolutions 
 (with the relativistic Doppler effect) are
 0.887~\kms\ and 0.532~\kms\ 
 for the \jthreetwo\ and \jfivefour\ transitions, respectively,
 for rest frequencies of 27294.289 MHz and 45490.3138 MHz 
 \citep{Mueller:01, Mueller:05}.
The data correlation process was repeated, adopting two different positions
 for the correlation center of \psrc, \ie the positions of the SW and NE images.
In J2000 coordinates, the pointing position for the observation was  
$(18^{\rm h}33^{\rm m}39\s 9, -21^\circ03'40\as5)$,
 and the adopted correlation positions were
 $(18^{\rm h}33^{\rm m}39\s 8816, -21^\circ03'40\as 6080)$ for the SW position and
 $(18^{\rm h}33^{\rm m}39\s 9316, -21^\circ03'39\as 8197)$ for the NE position.
 
\subsection{Calibration} 

The correlated data were calibrated using the NRAO Astronomical Image
 Processing System (AIPS; \citealt{Greisen:03}).
We did not detect absorption toward the NE component. 
Therefore, we analyzed only the data set with 
 the correlation center at the SW component position
 and only the 8MHz bands that covered 
 the absorption spectrum in each of the 14.5~GHz and 24.1~GHz bands.

We first corrected the interferometer delays and phases  
 for the effects of diurnal feed rotation (parallactic angle).
We also removed the off-source VLA station data
 by flagging the first minute of each scan.
The amplitudes of the interferometric data were corrected
 for the small biases ($\sim$ a few percent) of the data sampling thresholds
 among different stations,
 and then converted from correlation coefficients to flux density units 
 using the system temperature and antenna gain curve information.
  
We removed instrumental delay and phase offsets
 by fitting fringe patterns to the data of the calibrator \calib\ and
 corrected the data for these offsets. 
Bandpass amplitude and phase corrections were also determined
 from \calib\ and applied to \psrc.
We then performed phase calibration by fitting fringe patterns to the data of \psrc\
 on time scales of 20 seconds, using a model based on the image of \psrc\
 from a trial fringe fit.

A continuum spectral baseline for \psrc\
 was fitted based on channels 10 to 25 and 105 to 120
{ and subtracted from the visibility data, yielding
 a continuum-only data set and
 a baseline-subtracted ``line'' data set.}
The phase of the continuum data set was further calibrated
 using an iterative self-calibration process.
During the self-calibration process, we simultaneously CLEANed
 the SW and NE continuum components.
The NE component had a positional offset of $(\Delta \alpha \cos \delta, \Delta \delta)
 = (0\as6425, 0\as7276)$ relative to the SW component
 and was still in the field of view of the data correlated at the SW position.
The interferometer gain corrections from self-calibration 
 were then applied to the baseline-subtracted ``line'' data set of \psrc.

Before the spectrum fitting,
 we converted the frequency $\nu$ of the observed spectra into the velocity units $v$
 by using the relativistic Doppler effect,
 $\nu = \nu_0\sqrt{(1-v/c)/(1+v/c)}$, 
 where $c$ is the speed of light and $\nu_0$ the rest frequency of the transition.
We subtracted the expected velocity of
 168198.25 \kms\
 for the redshift of  $z=0.88582$,
 which defines the zero velocity of the obtained spectra.
 
\subsection{Rotational Temperature} \label{sect:trot}

From the measured absorption amplitude $\Delta S (v)$
 { (defined to be negative for absorption)}
 and continuum levels $S_{\rm c}$ of the two transitions,
 \hctnu\ and \hctnk,
 the optical depth $\tau (v)$ of each transition
{ is determined by

  \begin{eqnarray} \label{eqn:tau}
    \tau (v) = -\ln \left(1 + \frac{\Delta S(v)}{f_c S_c} \right),
  \end{eqnarray}
 where $f_c$ is the covering factor of the background continuum source
 by the absorbing cloud ($f_c \leq 1$).
If we assume the two transitions are caused by the same absorbing cloud,
  we can use}
 these values in the following equation 
 to determine the rotational temperature $T_{\rm rot}$ of the absorbing gas:
  \begin{eqnarray} \label{eqn:trot}
  T_{\rm rot} = \frac{-\Delta E_{42}}{k}\frac{1}{\ln\left( 
  \frac{33}{35}
  \frac{g_3}{g_5}
  \frac{\tau_{54}(v_0)}{\tau_{32}(v_0)} 
  \frac{ (1-{\rm e}^{-h\nu_{32}/kT_{\rm ex}}) }
           { (1-{\rm e}^{-h\nu_{54}/kT_{\rm ex}}) }\right)},
  \end{eqnarray}
 where $\Delta E_{42}$ is the energy difference between the $J=4$ and $J=2$ levels,
 $k$ the Boltzmann's constant 
 ($\Delta E_{42}/k = 3.06$~K;  \citealt{Mueller:01, Mueller:05}), 
 $g_{J} = 2 J + 1$ the statistical weight of the rotational level $J$,
 and $\nu_{ul}$ and $\tau_{ul} (v_0)$ are 
  the rest frequency and optical depth (at the line center velocity $v_0$),
  respectively, of the rotational absorption $J=u\leftarrow l$.
We assume equal excitation temperature $T_{\rm ex}$ for the two transitions \jthreetwo\ and \jfivefour\
(\ie local thermodynamical equilibrium)
 and also assume $\trot = \tex$ since collisional excitation is negligible for the diffuse interstellar gas. 
We solved Equation \ref{eqn:trot} for \trot\ numerically through iteration.
The equations used to derive Equation \ref{eqn:trot} are documented in the Appendix.

%==============================
 \section{RESULTS}\label{sect:results}

\subsection{Low Resolution Analysis}\label{sect:lowres}

\begin{figure}
\epsscale{1.1} 
\plottwo{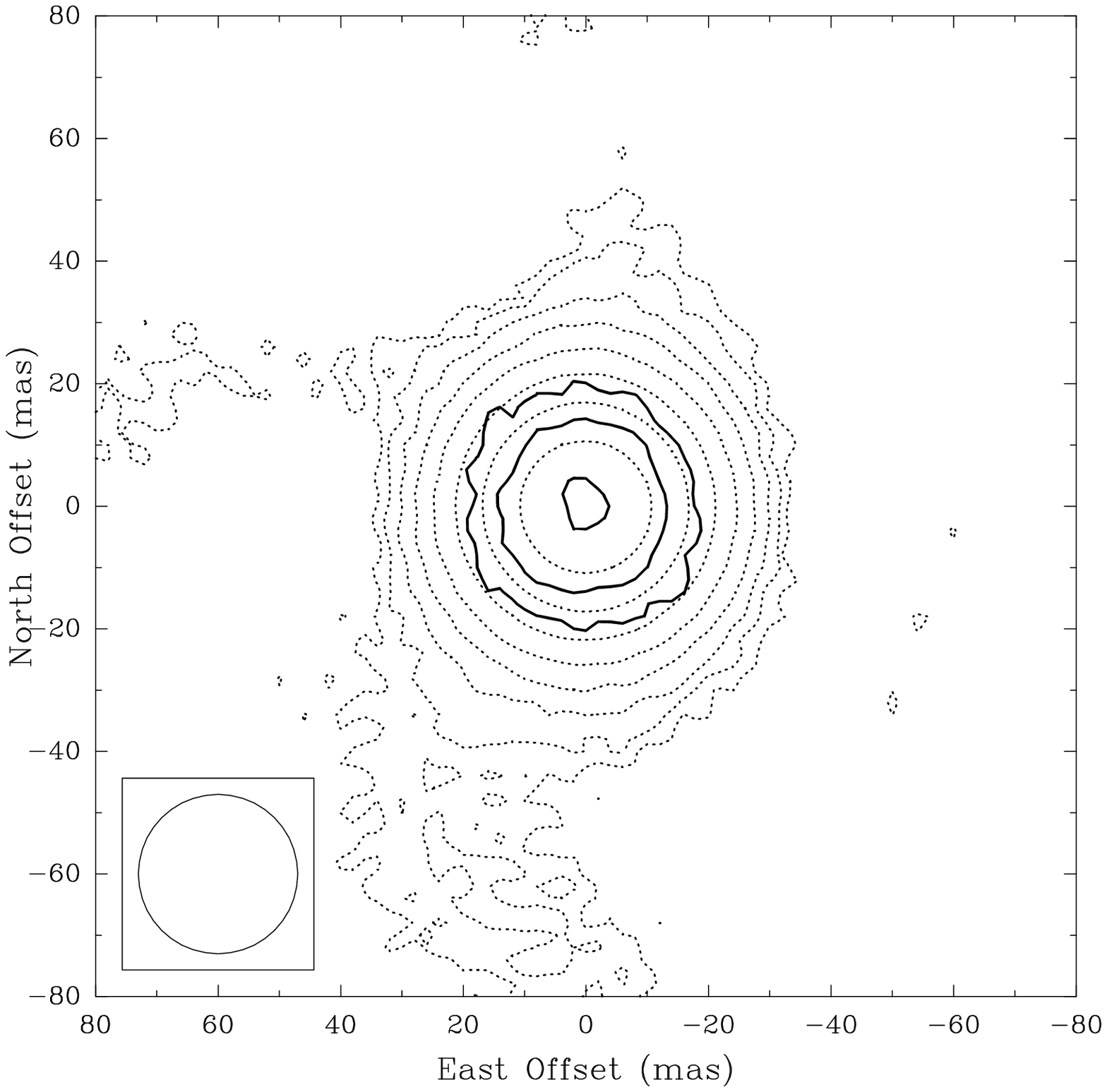} {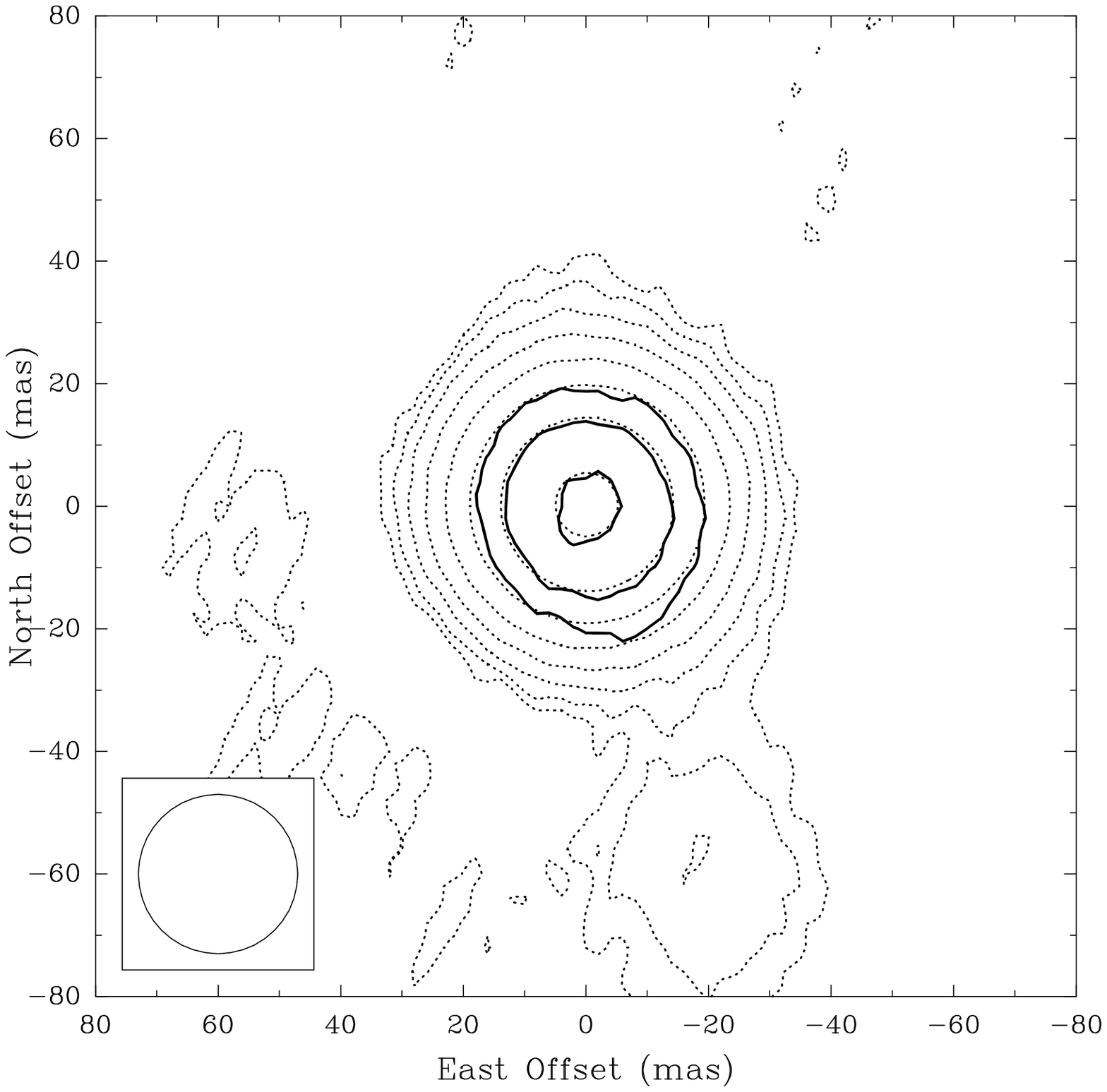}
\caption{
Absorption (solid) and continuum (dotted) maps with a { circular} beam of 26 mas size (FWHM)
  for the \hctnu\ line observed at 14.5~GHz (left) and the \hctnk\ line at 24.1~GHz (right).
The contours for both continuum and absorption are spaced by factors of 2, starting at 2$\sigma$,
 where $\sigma$ is the r.m.s.\ noise level of the { continuum} maps, $\sigma$ = 5 mJy/beam.
Note that the noise level in the continuum map is limited by the dynamic range
  and is comparable to the noise { level (4 mJy/beam)} in the absorption map.
{ The peak continuum levels (in the map center) are
 $S_{c,32} = 2.066$~Jy/beam for the \jthreetwo\ line and
 $S_{c,54} = 1.429$~Jy/beam for the \jfivefour\ line.}
} 
\label{fig:lowres}
\end{figure}

We first imaged the continuum and line absorption with low spatial resolution
 in order to simulate { the results from previous studies}.
We made image cubes at both 14.5~GHz and 24.1~GHz 
 with a circular restoring beam of 26 mas (FWHM);
 at this resolution, the { continuum} emission is essentially unresolved.
Figure~\ref{fig:lowres} displays these low resolution images;
{ the absorption distributions were obtained by plotting
 absorption depth, $|\Delta S|$, at each pixel of the image.}
We performed Gaussian fits to the observed spectra
 of the baseline-subtracted ``line''  image cubes
 and measured the absorption amplitudes of the
 \hctnu\ and \hctnk\ transitions { at each pixel of the image}.
{

We first solved for the three parameters, the amplitude, $\Delta S(v)$,
 the line width (FWHM), $\sigma_{v}$, and
  the central line velocity, $v_0$, of the absorption
  at the peak pixel of both continuum and absorption 
  at the center of the map.
We measured $v_{0,32} = -12.88 \pm 0.32$~\kms\ 
 and $\sigma_{v,32} = 5.94 \pm 0.77$~\kms\
 for the \jthreetwo\ line and  
 $v_{\rm 0,54} = -12.00 \pm 0.32$~\kms\
 and $\sigma_{v,54} = 8.49 \pm 0.81$~\kms\
 for the \jfivefour\ line.  
Averaging the results of the two frequency bands,
 we obtained the mean line velocity,
 $\bar{v}_0 = -12.44$~\kms,
  and the mean line width,
 $\bar{\sigma}_{v} = 7.21$~\kms.
We then repeated fitting at each pixel of the image
 with the line width and velocity fixed at the mean values, 
 $\bar{\sigma}_{v}$ and $\bar{v}_0$, solving for  
 the absorption amplitude, $\Delta S(v)$, only.
Table~\ref{tab:lowres} lists the measured values at
 the continuum peak pixel (in the map center) at each frequency,
and the \trot\ measurement  assuming 
 identical covering factors, $f_c$, for the two transitions.}
{ If we assume $f_c = 1$}, we measure  
{ $\trot = 5.6^{+2.5}_{\,-0.9}$~K, and varying $f_c$ gives similar results.}
This result is consistent with 
 the predicted value of $\tcmb = 5.14$~K
 and with previous measurements by 
 \citet[ $\trot \leq 6.0$~K]{Wiklind:96},
 \citet[$\trot = 4.5_{-0.6}^{+1.5}$~K]{Carilli:98}, 
 \citet{Henkel:09}
 and \citet{Muller:11}.
{ However, as shown in Section \ref{sect:tint}, if the absorption does not completely cover the continuum source, insufficient spatial resolution can yield biased rotational temperatures.}

%========================================
\subsection{High Resolution Analysis}\label{sect:hires}

\subsubsection{Imaging}\label{sect:imaging}
Next, we made images with the higher angular resolution afforded by 
 our long interferometer baselines.
From the line data,
 we made an image cube in each frequency channel
 with $128 \times 128$ pixels of 0.05 mas
 (\ie for a field of view of 6.4 mas $\times$ 6.4 mas).
We adopted identical circular restoring beams of 0.65 mas width (FWHM) for 
 the 14.5~GHz and 24.1~GHz bands.
(The dirty beam size was 
 1.45 mas $\times$ 0.46 mas with a position angle of $-1\d0$ at 14.5~GHz
 and 1.27 mas $\times$ 0.31 mas with a position angle of $-18\d3$ at 24.1 GHz.)

The line data were self-calibrated with the continuum at each frequency,
 and therefore the line and continuum maps are accurately aligned
 with a precision that is only limited by the uncertainty in the baseline calibration
 and by the thermal noise in the line image (due to the lower signal-to-noise ratio
 for absorption).
The bandpass phases are calibrated to better than 10 degrees, which translates 
 to a positional accuracy of $10/360$ of the fringe spacing, \ie $\sim0.03$~mas
 for a $\sim1$~mas fringe spacing. 
The positional uncertainty introduced by the thermal noise in the line image 
 can be estimated as half of the beam size $\theta_{\rm FWHM}$
 divided by the signal-to-noise ratio (SNR) of the image.
With $\theta_{\rm FWHM} =0.65$~mas and  SNR$\sim 6$ for absorption maps,
 this results in an accuracy of 0.05~mas.
{ For this case, relative positions measured in the line and continuum maps
 are accurate to better than 0.06 mas.}

Figure \ref{fig:map} shows the high resolution maps of the continuum
 emission of \psrc\ and the absorption in the foreground gas { at $z=0.89$.}
{ The continuum maps were obtained by imaging the
  continuum-only data set.
The absorption distribution in each frequency map
 was obtained by plotting absorption depth $|\Delta S|$ 
 measured at each pixel of the image (Section \ref{sect:spectfit}).}
Notably, we find that the absorption peak position in each frequency band,
{ in particular at 14.5~GHz,}
 is offset from the continuum peak position. 
{ The offsets are significantly larger than the uncertainty in the alignment
 between the line and continuum maps.} 
In the map of the 
 \jthreetwo\ transition observed at 14.5~GHz,
 the absorption peak lies at a positional offset of
$(\Delta \alpha \cos \delta, \Delta \delta)=(0.24\pm0.04, 0.56\pm0.05)$~mas 
 from the continuum peak, measured through fitting in the deepest absorption channel.
For the 
 \jfivefour\ transition observed at 24.1~GHz, 
 the measured offset is
 $(\Delta \alpha \cos \delta, \Delta \delta)=(0.04\pm0.01, 0.13\pm0.01)$~mas
  from the continuum peak to the absorption peak,
  fitted over the averaged map of the 10 channels ($\approx 5$~\kms\ width) in which the absorption was most prominent.
The presence of these offsets is a cause for concern that { the} rotational temperature,
 \trot, measured with lower angular resolution may be systematically biased
 (see Section \ref{sect:tint}). 

\subsubsection{Spectral fitting}\label{sect:spectfit}

Since the data of \psrc\ were self-calibrated,
 the absolute positions of both 14.5~GHz and 24.1~GHz maps are unknown,
 and hence we need to consider the 
 proper alignment of the two
{ frequency} maps before we measure \trot.
Were the continuum emission point-like, or { were its distribution}
 identical at the two frequencies, self-calibration
 would correctly align the two maps.
However, the continuum image of \psrc\ has resolved structure
 and the relative registration of the maps at the different 
 frequencies cannot be {\it a priori} determined.
Based on VLBI images of other sources, it is likely that
 the continuum emission of \psrc\ is elongated due to a faint one-sided jet 
 emanating from a self-absorbed core.  
Owing to optical depths varying
 with frequency for individual components, one expects a slight positional
 offset (along the jet axis) between the emission peaks at two frequencies.
(Note that since absorption depths are small compared to the continuum levels,
 the brightness peak positions of the continuum maps
 are not significantly affected by absorption.)

\begin{figure}
\epsscale{1.} 
\plottwo{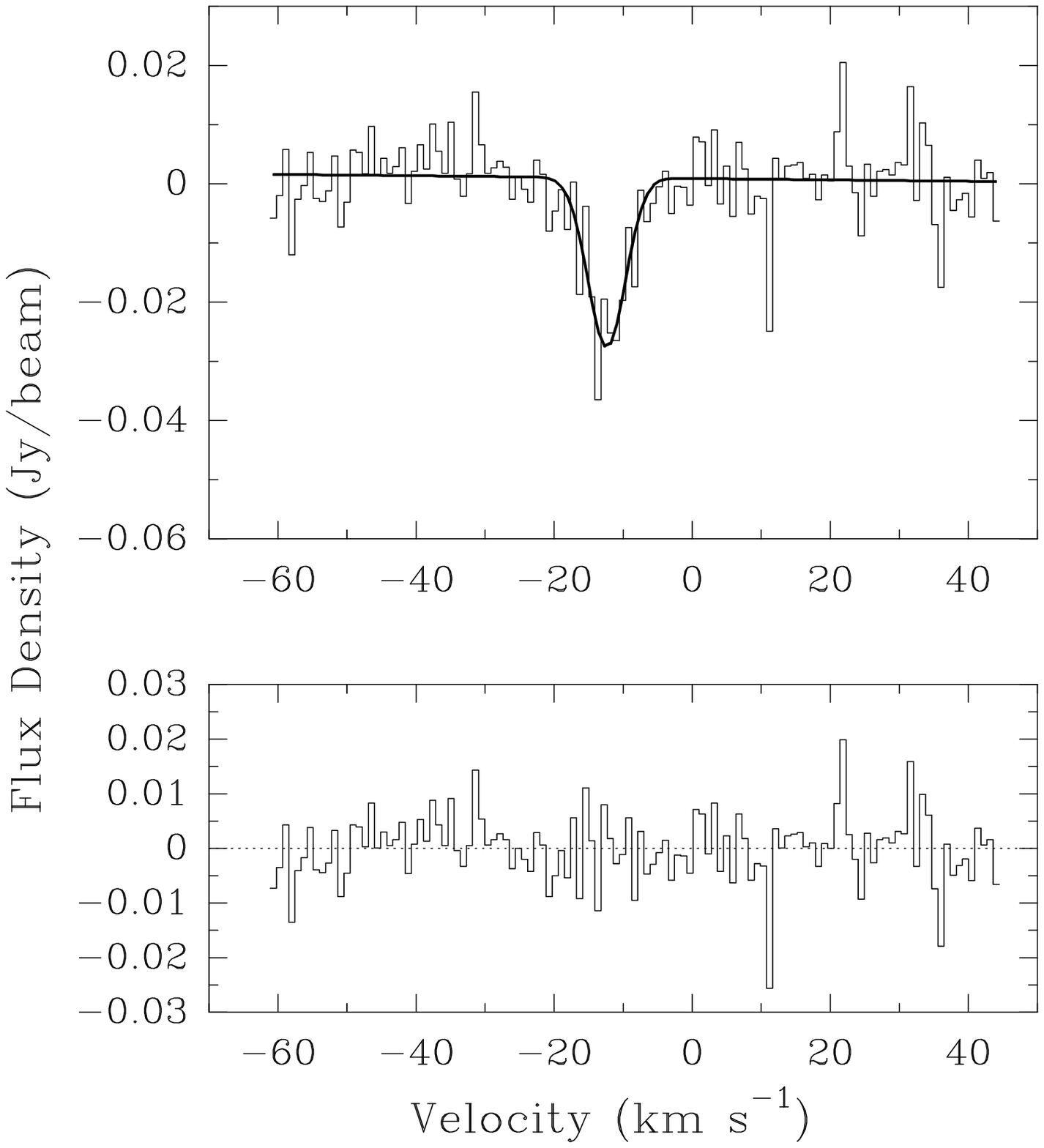} {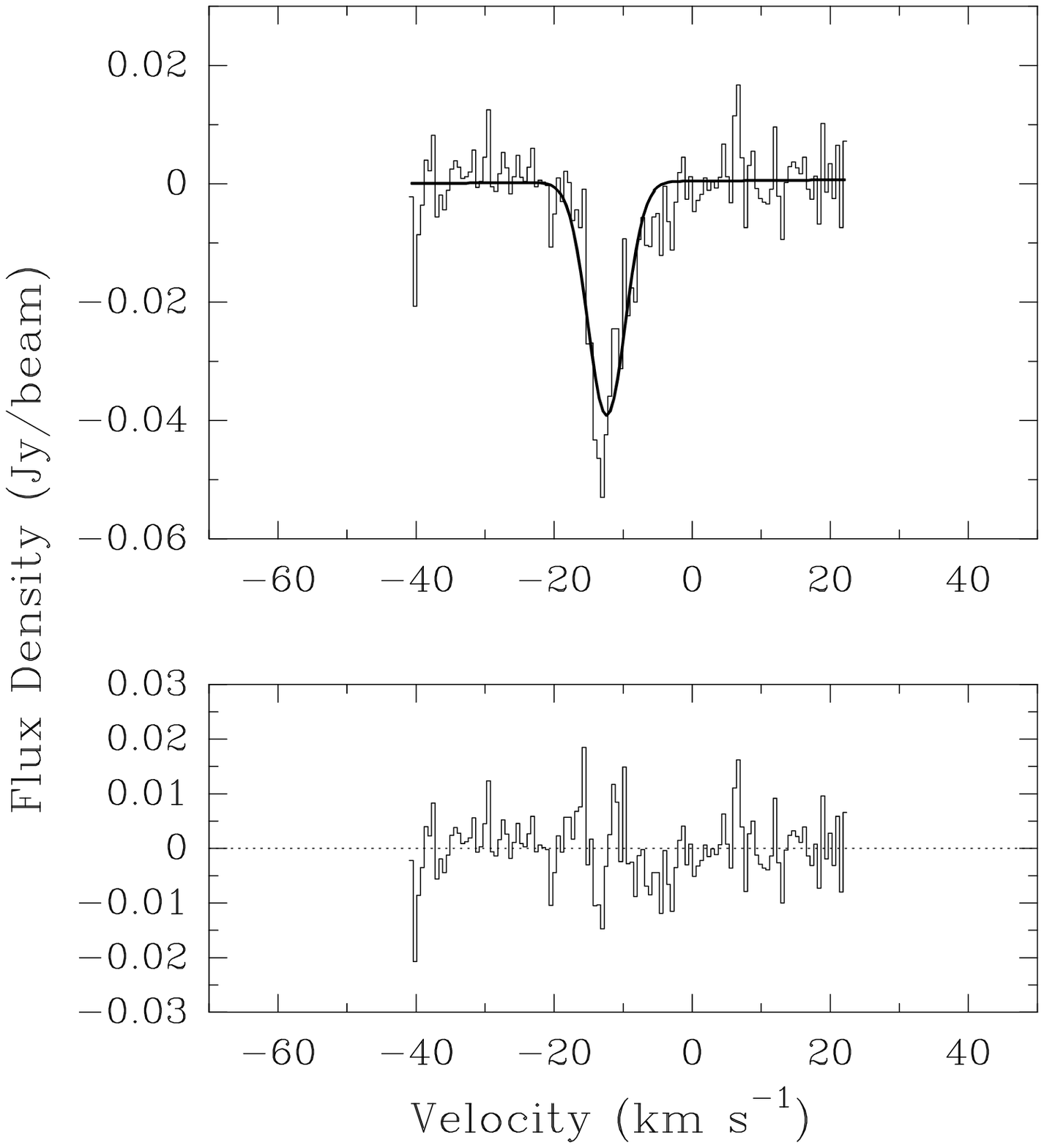}
\caption{
 { Baseline-subtracted} absorption spectra (upper panels) and residuals (bottom panels)
  of the \hctnu\ (left) and \hctnk\ (right) lines toward \psrc, 
  at the absorption peak positions.
{ At this position, the continuum levels are $S_{c,32} = 0.228$~Jy/beam for the \jthreetwo\ line
 and $S_{c,54}=0.816$~Jy/beam for the \jfivefour\ line.}
 The fitted spectra are shown by smooth curves in the upper panels.
 }
\label{fig:spect}
\end{figure}

We measured the absorption amplitude at different positions across the images,
 allowing evaluation of different alignments between the two images.
This was done in the following three steps using  
 least-squares fitting of Gaussian profiles to the spectra,
 { in the same manner as for the low resolution images.}
Firstly, we placed the measurement position for each frequency map
 at its absorption peak, and solved for the following
 three parameters: the amplitude $\Delta S(v)$
 the line width (FWHM) $\sigma_{v}$ and
  the central line velocity $v_0$ of the absorption.
{ We defined $\Delta S (v)$ as}
 $\Delta S (v) = S_{\rm line} (v) - S_{\rm c}$,
 where $S_{\rm line} (v)$ is the observed spectrum
 and $S_{\rm c}$ is the continuum level
  { (\ie $\Delta S (v) < 0$ for absorption).}
We then averaged the results of the two frequency bands 
 to determine the mean line width $\bar{\sigma}_{v}$
 and { the} mean velocity $\bar{v}_0$.
We obtained $\bar{v}_0 = -12.46$~\kms\ and
 $\bar{\sigma}_{v} = 6.38$~\kms\
 by averaging $v_{0,32} = -12.37 \pm 0.40$~\kms\ 
 and $\sigma_{v,32} = 7.27 \pm 0.98$~\kms\
 for the 
 \jthreetwo\ line and  
 $v_{\rm 0,54} = -12.54 \pm 0.17$~\kms\
 and $\sigma_{v,54} = 5.49 \pm 0.42$~\kms\
 for the 
 \jfivefour\ line.
Secondly, we repeated the fit at the same positions, 
 but fixed the line width and velocity at the mean values, 
 $\bar{\sigma}_{v}$ and $\bar{v}_0$, solving for  
 the absorption amplitude $\Delta S(v)$ only.
Figure~\ref{fig:spect} shows the result of this spectral fit.
The obtained values are 
 $\Delta S_{32} (v) = -28.6 \pm 2.8$~mJy/beam for the \jthreetwo\ line and
 $\Delta S_{54} (v) = -39.4 \pm 2.2$~mJy/beam for the \jfivefour\ line.
Finally, 
 using the same values of $\bar{v}_{\rm abs}$ and $\bar{\sigma}_{v}$,
 we measured the absorption amplitude $\Delta S(x,y,v)$ of all individual spectra
  at each pixel position $(x,y)$ across the two images.
{ The measured $\Delta S(x,y,v)$ is plotted as absorption distributions in Figure \ref{fig:map}.
} 

Previous VLBI observations of \psrc\ at 15 GHz and 43 GHz \citep{Garrett:97, Garrett:98, Jin:03}
 show complicated structure of both the SW and the NE images of the radio core with extended jet components
 on a sub-milliarcsecond scale.
In particular, multi-epoch 43 GHz maps by \citet{Garrett:97} have revealed
 that the core has
 remarkably complex substructure that changes rapidly with a timescale of weeks.
The complex morphology of the core substructure makes 
 it ambiguous to determine the jet direction.
We therefore determined the jet axis based on our continuum maps 
 to lie along the offset between the absorption and continuum peaks,
 and we shifted the position of the \trot\ measurement along this axis
 to allow for the possibility of misalignment of the two maps.

\subsubsection{\trot\ measurement}

\begin{figure}
\epsscale{1.1} 
%\plottwo{f1a.eps} {f1b.eps}
\plottwo{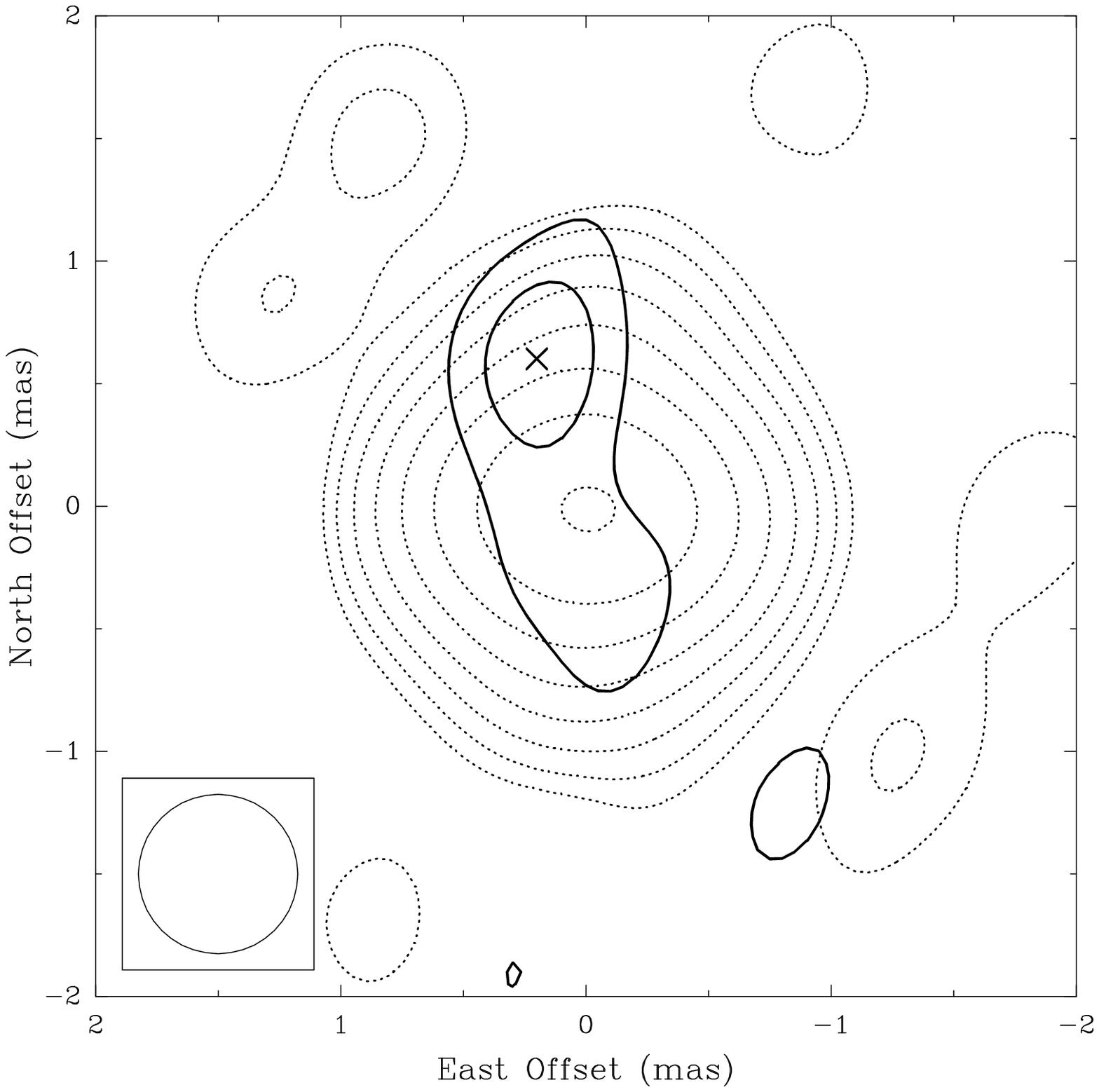} {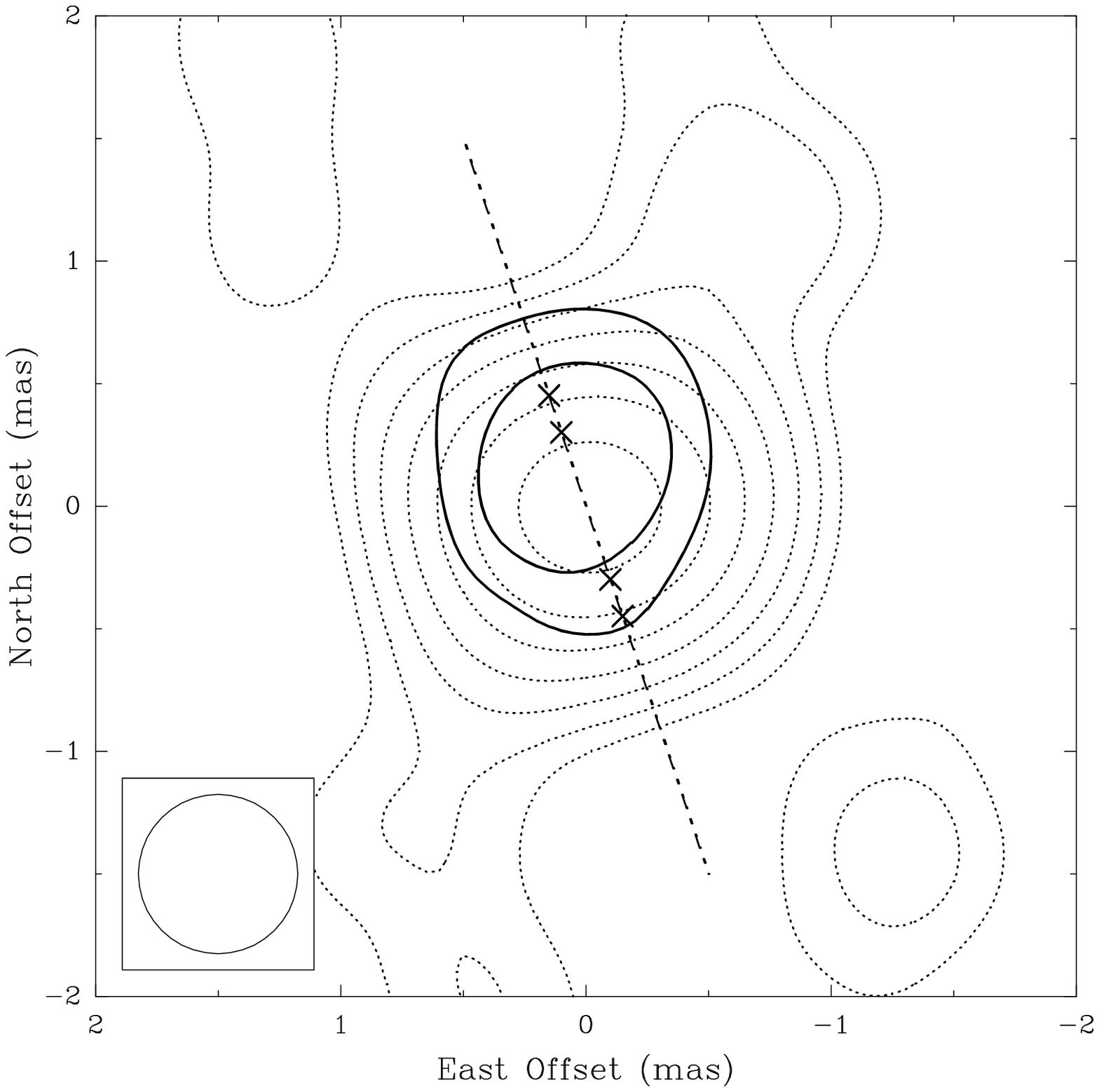}
\caption{
Same as Figure 1, but with a beam size (FWHM) of 0.65 mas, with the same r.m.s.\ noise level
 of the continuum maps, $\sigma$ = 5 mJy/beam (adopted for both absorption and continuum contours).
{ The peak continuum levels (in the map center) are $S_{c,32} = 1.340$~Jy/beam for the \jthreetwo\ line
 and $S_{c,54} = 0.949$~Jy/beam for the \jfivefour\ line. }
 Measurement positions for \trot\ are marked by crosses { (corresponding to Table~\ref{tab:hires})}, along the dash-dotted line
  connecting the continuum and absorption peaks (see text).}
\label{fig:map}
\end{figure}

For a given alignment of the two images, \trot\ can be measured at any pixel position,
 provided that both absorption and background continuum emission have sufficient signal-to-noise
ratios.
However, since the detected absorption was weak for the \jthreetwo\ line, 
 we fixed the position for measuring its optical depth, $\tau_{32}$,
 at its absorption peak position in the 14.5~GHz map,
 maximizing the signal-to-noise ratio of the measurement.
We shifted the measurement point of $\tau_{54}$ along the line shown as a dash-dotted line in
 Figure~\ref{fig:map}.
{ Both continuum and absorption levels were measured at a single pixel, and we performed
 \trot\ measurements using three different covering factors ($f_c =  1, 0.5, 0.2$),
 assumed to be identical for the two transitions.}

Of course only a very limited range of shifts make physical sense, and we required
 the continuum emission to reflect realistic synchrotron spectral indices for all pixels
 with firm detections.
Our selection criteria required spectral indices, $-1 < \alpha < 2$
{  (defined by the power-law spectrum $S_\nu \propto \nu^{\alpha}$), }
 which yielded four possible alignments shown by crosses in the 24.1 GHz map in Figure~\ref{fig:map}. 
{ Table~\ref{tab:hires} lists the measured values from these measurement points
 (as well as from excluded measurement positions with spectral indices outside of $-1 < \alpha < 2$).}
These alignments yield \trot\ between { 1.1} and { 2.5} K, 
 which is significantly lower than 
 the theoretically-predicted value for
 the cosmic microwave background, $\tcmb = 5.14$~K at $z = 0.88582$.
{ (The three excluded measurement positions in Table \ref{tab:hires} near the map center, 
 \ie $(\vert \Delta x \vert, \vert \Delta y \vert) < (0.1, 0.3)$ mas, 
yield spectral indices, $\alpha$, between 2.5 and 2.8, and \trot\ values between 1.1 and 1.6 K.)}
We note, as \citet{Henkel:09} pointed out, that due to the small rotational constant of \hctn,
 the measured excitation temperature may be susceptible to additional contributions
 from collisional excitation from (even moderate densities of) H$_2$ gas
 or local radiative excitation.
While this could explain a higher { value} of \tcmb\ (as in \citealt{Henkel:09}),
 it cannot explain why our measurement is lower than the radiation temperature.
We discuss the dominant biases in the present result in Section \ref{sect:tint}.
{
By exploiting \trot\ measurements using Equation \ref{eqn:trot},
 we assume that the physical properties of the gas is the same for the two transitions
 (\ie the both absorption lines come from the same absorbing cloud).
We will also discuss this assumption in Section \ref{sect:tint}. 
}

\section{BIAS OF TEMPERATURE MEASUREMENT}\label{sect:tint}

{ Our high-resolution measurement yields \trot\ between { 1.1} and { 2.5}~K,
 which is significantly lower than the expected value of $\tcmb=5.14$~K at $z = 0.89$. 
In this section, we discuss several possible explanations for this discrepancy.
}

\subsection{Contribution from non-absorbed continuum emission}

Our VLBI maps of \psrc\ have revealed
 that the absorption peak positions of the \hctnu\ and \hctnk\ lines
 are offset from the continuum peak of the SW core.
Because these two lines were observed contemporaneously, temporal variations can be 
 excluded, { and these offsets} indicate that the absorbing cloud
 of the foreground galaxy is covering the { core-jet emission from} \psrc\ { only partially, and not entirely}.
 
The synchrotron radiation from 
 the extended components
 of AGNs such as jets
 is optically thin at cm wavelengths, 
 with a spectral index $\alpha \approx -0.7$ to $-1.0$ 
 (defined as $S_\nu \propto \nu^{\alpha}$), 
 whereas the compact core is generally optically thick 
 and synchrotron self-absorption at low frequencies 
 inverts the spectrum toward $\alpha > 0$ 
 (\eg \citealt{Schneider:06, Rybicki:79}).
Since the core is very compact and unresolved on milliarcsecond scales,
 our continuum measurements likely include a blend of jet and core emission.

Blending {\it non-absorbed} continuum emission with $\alpha > 0$
 within the beam causes the ratio of the optical depths
\begin{displaymath}
  \frac{\tau_{54} (v_0)}{\tau_{32} (v_0)} \approx
  \frac{\Delta S_{54}(v_0)}{\Delta S_{32} (v_0)}\frac{S_{c,32}}{S_{c,54}}
\end{displaymath}
 to decrease from the true value, 
 because the continuum level $S_{c,54}$ 
 relative to $S_{c,32}$ ($\nu_{54} > \nu_{32}$) is increased (due to $\alpha > 0$), 
 while the absorption terms remain unchanged.
This biases the value of \trot\ given by Equation \ref{eqn:trot} to be lower than
 the true \trot\ value as we observe.

In case of our low-resolution measurement (Section~\ref{sect:lowres}), 
 we are likely more sensitive to extended, low surface brightness emission
 that could have a significant contribution of jet emission that is also not absorbed,
{ resulting in our measured spectral index of $\alpha = -0.7 \pm 0.6$ (Table~\ref{tab:lowres}).} 
The addition of non-absorbed continuum emission from extended components 
 with $\alpha < 0$ can falsely increase the measured \trot.
Of course, the rough match between the measured and expected \trot\ values 
 would then have to be coincidental. 

\subsection{Different absorbing gas}

{
In our \trot\ measurements, we assumed identical physical conditions for
 { the gas giving rise to} the two transitions.
Our high resolution maps (Figure~\ref{fig:map})
 show that the absorption peak of the \jfivefour\ line is located more closely to
 the continuum peak than the \jthreetwo\ absorption peak (Section~\ref{sect:imaging}),
 and also the width of the absorption line in Figure~\ref{fig:spect} is narrower 
 for the \jfivefour\ transition than for the \jthreetwo\ transition,
 which could be interpreted as the two absorption lines originating in different gas clouds.
If this were the case, no relation between the optical depths of the two absorption lines could be assumed,
 and therefore any value for \trot\ could have resulted from the measurement. 
Theoretically, it is possible to detect only \jthreetwo\ absorption from a given cloud
 where the \jfivefour\ line is very weak at low temperature, but 
 when the \jfivefour\ line is detected, the \jthreetwo\ line must also be detected
 except under highly unusual circumstances.
Therefore, it is very unlikely that we see two different absorbing clouds. 
}

\subsection{Core-shift of \psrc\ and misalignment at two frequencies}
{ The observed difference in offset between continuum and absorption peaks at the two frequencies
 could also be caused
 by different apparent positions of the core at the two frequencies, rather than differences in the foreground gas.
The radio core positions of AGNs are known to shift with frequency
 due to frequency-dependent optical depths of the jets \citep{Blandford:79}.
Such a core-shift effect has been observed, for example, recently by \citet{Hada:11} in M87, but 
 with only 0.1 mas shift between frequencies of 15.4 and 23.8 GHz, comparable
 to our observing frequencies.
This is well within our resolution, and considering the larger distance to \psrc,
 this effect should be negligible in our data.
}

%=========================================================================
\section{CONCLUSIONS}\label{sect:concl}
Our VLBI observations of 
 \psrc~SW have identified offsets between the absorption (\hctn\ \jthreetwo\ and \jfivefour\ lines)
 and continuum peaks (all observed contemporaneously). 
{ These offsets indicate} that absorption in the foreground galaxy covers
 the { core-jet emission from} \psrc\ { only partially}.
We have shown that insufficient spatial resolution { can cause} 
unresolved, non-absorbed continuum emission
 to contaminate \trot\ measurements, 
 and that all previous \trot\ measurements could be affected by this bias
 { (although measurements at millimeter wavelengths may be less affected by this
 bias because the jet component becomes less pronounced)}.
Our high-resolution measurement yields \trot\ between { 1.1} and { 2.5}~K,
 which is significantly lower than the expected value of $\tcmb=5.14$~K from
 the cosmic microwave background at $z = 0.88582$.
The true rotational temperature of the absorbing gas is likely higher than
 { this} measured \trot\ value, because of blended-in
 core-jet components with spectral index $\alpha >0$, which biases
 the measurement toward a lower temperature. 
Our result reinforces the importance of high-resolution observations of \psrc\
 to map the absorbing regions in detail, in order to reliably measure the
 rotational temperature and the cosmic microwave background.
More observations with higher sensitivity to measure absorption depths
 as well as an absolute-position measurement for correct map alignment 
 will further improve the accuracy of the measurement.

\acknowledgements
MS acknowledges financial support from a JSPS Postdoctoral Fellowship for Research Abroad.
The authors are grateful to Christian Henkel and Arnaud Belloche for careful reading of the manuscript and 
 to Joris Verbiest for discussions and contributions to the manuscript.
We thank the referee Professor Francoise Combes for insightful comments and suggestions
 that improved the paper.

{\it Facilities:} \facility{VLBA}

\section*{APPENDIX}
Our VLBA observations of the southwest (SW) core of \psrc\ were
 used to estimate the rotational temperature \trot\ of the foreground gas
 at $z = 0.88582$ by measuring absorption depths of \hctn\ in
 the \jthreetwo\ and \jfivefour\ lines.
In this section, we derive Equation \ref{eqn:trot} 
 that was used 
 to measure the rotational temperature of the gas
 from the observed absorption depths.

The ratio of the volume densities $n$ of \hctn\ molecules in the lower states
  of the two transitions are given by:
  \begin{displaymath}
  \frac{n_4}{n_2} = \frac{g_4}{g_2} \exp
  \left( \frac{-\Delta E_{42}}{kT_{\rm rot}} \right) ,
  \end{displaymath}
 where $g_J = 2 J + 1$ are the statistical weights of the levels,
 $k$ is the Boltzmann's constant, and
 $\Delta E_{42}$ is the energy difference between the $J=4$ and $J=2$ levels
 ($\Delta E_{42}/k = 3.06$~K; \citealt{Mueller:01, Mueller:05}).
Solving the above equation for the rotational temperature yields:
 \begin{displaymath}
     T_{\rm rot} = \frac{-\Delta E_{42}}{k}\frac{1}
     {\ln \left( \frac{n_4}{n_2} \frac{g_2}{g_4} \right)}.
 \end{displaymath} 
We now evaluate the density ratio $n_4 / n_2$
 in terms of the observed optical depths $\tau_{32} (\nu)$ and $\tau_{54} (\nu)$
 of the \jthreetwo\ and \jfivefour\ lines, respectively. 
For each transition, the optical depth $\tau (\nu)$ is given by:
 \begin{displaymath}
    \tau(\nu) = \int_0^L \kappa (\nu) {\rm d}l = \kappa (\nu) L,
 \end{displaymath}
 where $\kappa (\nu)$ is the absorption coefficient, assumed uniform,
 and $L$ is the total path length through the absorbing cloud.
The absorption coefficient $\kappa (\nu)$ for a transition 
 from the upper level $u$ to the lower level $l$ is given by:
  \begin{displaymath}
     \kappa (\nu) = \frac{1}{8\pi}\frac{c^2}{\nu_{ul}^2}
     \frac{g_u}{g_l}n_l A_{ul} \left(
        1-{\rm e}^{-h\nu_{ul}/kT_{\rm ex}} \right)
      \phi (\nu),
  \end{displaymath}
 where $c$ is the speed of light,
 $\nu_{ul}$ is the transition frequency,
 $A_{ul}$ is the Einstein coefficient of spontaneous emission, 
 $h$ is the Planck's constant,
 $T_{\rm ex}$ is the excitation temperature,
 and $\phi (\nu)$ is the normalized line profile,
 which can be represented for any transition as:
 \begin{displaymath}
   \phi (\nu) = \frac{1}{\sqrt{2\pi}\sigma{'}_{\nu}} \,{\rm e}^{-(\nu-\nu_0)^2 / 2 \sigma{'}_{\nu}^2},
 \end{displaymath}
  where $\sigma{'}_{\nu}$ is the frequency half-width at e$^{-0.5}$ of the peak
  and $\nu_0$ is the line center frequency.
This can be rewritten in terms of Doppler velocity as:
 \begin{displaymath}
   \phi (v) = \frac{c}{\nu_0} \frac{1}{\sqrt{2\pi}\sigma{'}_{v}} \,{\rm e}^{-(v-v_0)^2 / 2 \sigma{'}_{v}^2},
 \end{displaymath}
 where $\sigma{'}_{v}$ is the velocity half-width at e$^{-0.5}$ of the peak,
 which is related to the full-width at half-maximum (FWHM) $\sigma_{v}$ by $\sigma_{v} = \sqrt {8 \ln 2} \,\sigma{'}_{v}$,
 and $v_0$ is the line center velocity.
Assuming the line profiles of the two transitions are identical in Doppler velocity
 (\ie $\sigma{'}_{v32} = \sigma{'}_{v54}$), 
 then at line center ($v=v_0$ and $\nu = \nu_0$), the ratio of the line profiles is given by the reciprocal ratio of the rest frequencies:
 \begin{displaymath}
   \frac{\phi_{32} (v_0)}{\phi_{54} (v_0)} = \frac{\nu_{54}}{\nu_{32}}.
 \end{displaymath} 
Therefore, taking the ratio of the optical depths
 $\tau_{32} (\nu_0)$ and $\tau_{54} (\nu_0)$ at line center
 and assuming equal path lengths
 and excitation temperatures for the two transitions yields the following relation:
  \begin{displaymath}
    \frac{\tau_{32} (\nu_0)}{\tau_{54} (\nu_0)} = \left( \frac{\nu_{54}}{\nu_{32}}\right)^3
    \frac{g_3}{g_5} \frac{n_2 g_4}{n_4 g_2} \frac{A_{32}}{A_{54}}
     \left( \frac{1-{\rm e}^{-h\nu_{32}/kT_{\rm ex}}}
                        {1-{\rm e}^{-h\nu_{54}/kT_{\rm ex}}} \right).
 \end{displaymath}
Solving this equation for the density ratio term needed to estimate a rotational temperature, we obtain:
  \begin{displaymath}
    \frac{n_4 g_2}{n_2 g_4} =
    \frac{\tau_{54} (\nu_0)}{\tau_{32} (\nu_0)}
    \left( \frac{\nu_{54}}{\nu_{32}}\right)^3
    \frac{g_3}{g_5}  \frac{A_{32}}{A_{54}}
    \left( \frac{1-{\rm e}^{-h\nu_{32}/kT_{\rm ex}}}
                       {1-{\rm e}^{-h\nu_{54}/kT_{\rm ex}}} \right).
 \end{displaymath}
For $J \rightarrow J+1$ absorption transitions, the Einstein A coefficients are given by:
 \begin{displaymath}
 A_{J+1, J} = \frac{64 \pi^4}{3hc^3} \, \mu^2 \nu^3_{J+1,J} \, \frac{J+1}{2J+3},
 \end{displaymath}
 where $\mu$ is the permanent electric dipole moment of the molecule,
 and thus, the ratio of the Einstein A coefficients (for $J= 2$ and $J=4$) are given by:
 \begin{displaymath}
   \frac{A_{32}}{A_{54}} = \frac{3/7}{5/11} \left( \frac{\nu_{32}}{\nu_{54}}\right)^3 = \frac {33}{35} \left( \frac{\nu_{32}}{\nu_{54}}\right)^3.
 \end{displaymath}
Therefore, we obtain the desired rotational temperature \trot\
 in terms of observed absorption optical depths
 of the \hctn\ \jthreetwo\ and \jfivefour\ lines as:
  \begin{displaymath} 
  T_{\rm rot} = \frac{-\Delta E_{42}}{k}\frac{1}{\ln\left( 
  \frac{33}{35}
  \frac{g_3}{g_5}
  \frac{\tau_{54}(v_0)}{\tau_{32}(v_0)} 
  \frac{ 1-{\rm e}^{-h\nu_{32}/kT_{\rm ex}} }
           { 1-{\rm e}^{-h\nu_{54}/kT_{\rm ex}} }\right)}.
  \end{displaymath}

%----------------------------

%============================================================
%%%  FIGURES start here %%%%%%%%%%%%%%%

%\appendix

\begin{deluxetable}{lllcccc}
\tablecolumns{7} \tablewidth{0pc} 
\tablecaption{\trot\ measurement from the low-resolution analysis}\tablehead {
  \colhead{Transition} &
  \colhead{$S_{c}$} & \colhead{$\Delta S$} &
  \colhead{$\alpha$} & \colhead{$f_c$} & \colhead{$\tau$} & 
  \colhead{\trot} 
\\
  \colhead{} &  \colhead{(Jy/beam)} & \colhead{(Jy/beam)} &
  \colhead{  }  & \colhead{} & \colhead{} & \colhead{(K)}
            }
\startdata
 \jthreetwo & 2.069 (0.005) & $-$0.048 (0.004) &  ---                      &  1.0 & 0.023 (0.002)  &  --- \\
                     &                          &                                 &                            & 0.5 &  0.047 (0.005)  &  \\
                     &                          &                                 &                            & 0.2 &  0.123 (0.012) &  \\                     
 \jfivefour   & 1.429 (0.005) & $-$0.049 (0.003) & $-$0.7 (0.6)  & 1.0 & 0.035 (0.003)  & 5.6$^{+2.5}_{\,-0.9}$ \\
                     &                          &                                 &                            & 0.5 & 0.070 (0.005)  & 5.6$^{+2.9}_{\,-0.9}$\\
                     &                          &                                 &                            & 0.2 & 0.186 (0.015)  & 5.9$^{+5.9}_{\,-1.0}$\\
 \enddata
\tablecomments{Measured values from Figure \ref{fig:lowres}, at the continuum peak pixel in the map center.  
$S_{c}$ is the continuum level, $\Delta S$ is the absorption level,
 $\tau$ is the optical depth of each transition,
 and $\alpha$ is the spectral index derived from the continuum levels at the two frequencies.
The uncertainties are listed in parentheses.
The uncertainty of the spectral index, 0.6, was derived
 assuming a systematic error of 20\% in flux scale calibration at each frequency band. 
 (Note the flux-scale error does not affect \trot\ values.) 
The rotational temperature \trot\ was measured from the two transitions,
 assuming identical covering factors $f_c$.
} 
\label{tab:lowres}
\end{deluxetable}

\begin{deluxetable}{lrrccrccc}
\tablecolumns{9} \tablewidth{0pc} 
\tablecaption{\trot\ measurement from the high-resolution analysis}
\tablehead {
  \colhead{Transition} &
  \colhead{$\Delta x$} & \colhead{$\Delta y$} &
  \colhead{$S_{c}$} & \colhead{$\Delta S$} &
  \colhead{$\alpha$} & \colhead{$f_c$} & \colhead{$\tau$} & 
  \colhead{\trot} 
\\
  \colhead{} & 
  \colhead{(mas)}      & \colhead{(mas)} &
  \colhead{(Jy/beam)} & \colhead{(Jy/beam)} &
  \colhead{  }  & \colhead{} & \colhead{} & \colhead{(K)}
            }
\startdata
\jthreetwo & 0.30      & 0.60        & 0.229 & $-0.029$ & ---               & 1.0 & 0.13 (0.02)  &--- \\
             &               &                 &             &                  &                     & 0.5 & 0.29 (0.05)  & \\
             &               &                 &             &                  &                     & 0.2 & 1.01 (0.26)  & \\
\jfivefour & 0.20    &  0.60       & 0.111 & $-0.017$ & $-1.4$     & 1.0 & 0.17 (0.04)  & 4.3$^{+70.6}_{\,-1.2}$\\
            &               &                 &            &                   &                      & 0.5 & 0.38 (0.11) & 4.5$^{+77.9}_{\,-1.4}$\\
            &               &                 &            &                   &                     & 0.2 & 1.35 (0.53) & 6.0$^{+821.0}_{\,-2.7}$\\
           & 0.15   &  0.45       & 0.274 & $-0.026$    &  \bf 0.4              & 1.0  & 0.10 (0.02)  & 2.5$^{+0.8}_{\,-0.4}$\\
           &               &                 &            &                   &                     & 0.5 & 0.21 (0.04)  & 2.5$^{+1.2}_{\,-0.4}$\\
           &               &                 &            &                   &                     & 0.2 & 0.67 (0.15)  & 2.3$^{+1.2}_{\,-0.4}$\\
           & 0.10      &  0.30       & 0.537 & $-0.035$ &   \bf 1.7              & 1.0  & 0.07 (0.01) & 1.9$^{+0.3}_{\,-0.2}$ \\
           &               &                 &            &                   &                     & 0.5 & 0.14 (0.02) & 1.9$^{+0.3}_{\,-0.2}$\\
           &               &                 &            &                   &                     & 0.2 & 0.40 (0.06) & 1.7$^{+0.4}_{\,-0.2}$\\
           & 0.05      &  0.15       & 0.816 & $-0.039$ &  2.5             & 1.0  & 0.05 (0.01) &  1.6$^{+0.2}_{\,-0.1}$\\
            &               &                 &            &                   &                      & 0.5 & 0.10 (0.01) &  1.5$^{+0.2}_{\,-0.1}$\\
           &               &                 &            &                   &                      & 0.2 & 0.28 (0.03) &  1.4$^{+0.2}_{\,-0.2}$\\
           & 0.00      &  0.00       & 0.949 & $-0.035$ &  2.8            & 1.0 & 0.04 (0.01) &  1.4$^{+0.2}_{\,-0.1}$\\
           &               &                 &            &                   &                      & 0.5 & 0.08 (0.01) &  1.4$^{+0.2}_{\,-0.1}$\\
           &               &                 &            &                   &                      & 0.2 & 0.21 (0.03) &  1.2$^{+0.2}_{\,-0.1}$\\
         & $-0.05$  & $-0.15$ & 0.839 & $-0.025$ & 2.5             & 1.0 & 0.03 (0.01) & 1.3$^{+0.1}_{\,-0.1}$\\
         &               &                 &             &                   &                      & 0.5 & 0.06 (0.01) & 1.2$^{+0.1}_{\,-0.1}$\\
         &               &                 &            &                   &                       & 0.2 & 0.17 (0.03) &  1.1$^{+0.2}_{\,-0.1}$\\
       & $-0.10$ & $-0.30$  & 0.563 & $-0.016$ &  \bf 1.8             & 1.0 & 0.03 (0.01) & 1.2$^{+0.2}_{\,-0.1}$ \\
         &               &                 &            &                   &                     & 0.5 & 0.06 (0.02) & 1.2$^{+0.2}_{\,-0.1}$\\
         &               &                 &            &                   &                     & 0.2 & 0.15 (0.04) & 1.1$^{+0.3}_{\,-0.2}$\\
        & $-0.15$ & $-0.45$  & 0.289 & $-0.011$ &  \bf 0.5            & 1.0 & 0.04 (0.01) & 1.4$^{+0.3}_{\,-0.3}$ \\
        &               &                 &            &                   &                     & 0.5 & 0.08 (0.03)  & 1.4$^{+0.3}_{\,-0.3}$\\
        &               &                 &            &                   &                     & 0.2 & 0.21 (0.09)  & 1.2$^{+0.3}_{\,-0.2}$\\
        & $-0.20$ & $-0.60$  & 0.115 & $-0.009$ & $-1.3$       & 1.0 & 0.08 (0.04) & 2.1$^{+1.2}_{\,-0.6}$ \\
        &               &                 &            &                   &                     & 0.5 & 0.17 (0.08)  & 2.0$^{+17.8}_{\,-0.6}$\\
        &               &                 &            &                   &                     & 0.2 & 0.52 (0.30)  & 1.8$^{+60.6}_{\,-0.5}$\\
\enddata        
\tablecomments{
Measured values from Figure~\ref{fig:map}.  
$\Delta x$ and $\Delta y$ show the positions (measured in 0.05 mas pixel size) of each measurement 
 relative to the continuum peak position in the map center at each frequency.
$S_{c}$ is the continuum level, $\Delta S$ is the absorption level,
 $\tau$ is the optical depth of each transition,
  and $\alpha$ is the spectral index derived from the continuum levels at the two frequencies,
  shown in bold font if it satisfies our selection criteria of $-1 < \alpha < 2$.
The uncertainties (or noise levels) of the continuum $S_{c}$ and the absorption level $\Delta S$
 are 0.005~Jy/beam and 0.004~mJy/beam, respectively. 
The uncertainty of each spectral index is 0.6,
 assuming a systematic error of 20\% in flux scale calibration at each frequency band. 
 (Note the flux-scale error does not affect \trot\ values, since \trot\ is measured using the relative ratios of the absorption and the continuum levels at each band.)
We measured the rotational temperature \trot\ from the two transitions, 
 assuming identical covering factors $f_c$.
Measurements for the \jfivefour\ transition were made along the axis shown in Figure~\ref{fig:map}.
Note that excluded measurement positions with spectral indices outside of $-1 < \alpha < 2$ are also listed for comparison.
} 
\label{tab:hires}
\end{deluxetable}

\end{document}